\documentclass[runningheads]{llncs}
\usepackage[T1]{fontenc}
\usepackage{graphicx}
\usepackage{makecell}
\usepackage{booktabs}
\usepackage{hyperref}
\begin{document}
\title{EMERS: Energy Meter for\\ Recommender Systems}
%
%
\author{Lukas Wegmeth\inst{1}\orcidID{0000-0001-8848-9434} \and
Tobias Vente\inst{1}\orcidID{0009-0003-8881-2379} \and\\
Alan Said\inst{2}\orcidID{0000-0002-2929-0529} \and
Joeran Beel\inst{1}\orcidID{0000-0002-4537-5573}}

\institute{University of Siegen, Germany \and University of Gothenburg, Sweden}
%
%
%
\maketitle              
\begin{abstract}
Due to recent advancements in machine learning, recommender systems use increasingly more energy for training, evaluation, and deployment.
However, the recommender systems community often does not report the energy consumption of their experiments.
In today's research landscape, no tools exist to easily measure the energy consumption of recommender systems experiments.
To bridge this gap, we introduce EMERS, the first software library that simplifies measuring, monitoring, recording, and sharing the energy consumption of recommender systems experiments.
EMERS measures energy consumption with smart power plugs and offers a user interface to monitor and compare the energy consumption of recommender systems experiments.
Thereby, EMERS improves sustainability awareness and simplifies self-reporting energy consumption for recommender systems practitioners and researchers.

\end{abstract}
\section{Introduction}
Recommender systems research seldom discusses and reports energy consumption for executing experiments~\cite{10.1145/3604915.3608840,10.1145/3640457.3688074}.
This is exacerbated by missing author guidelines\footnote{\url{https://recsys.acm.org/recsys24/call/}} and tools~\cite{10.1145/3604915.3608840,10.1145/3640457.3688074} for measuring and reporting energy consumption.
We identify two significant reasons why measuring and reporting the energy consumption of recommender systems experiments is necessary.

The first reason is to understand the environmental impact of recommender systems experiments to facilitate decision-making toward green solutions.
Scientists agree that climate change is man-made~\cite{Lynas2021,Cook2016}.
Electricity generation creates greenhouse gases that accelerate climate change\footnote{\url{https://www.iea.org/reports/co2-emissions-in-2023}}.
According to optimistic forecasting, computing will contribute to global energy usage by more than 7\% in 2030~\cite{9407142}.
However, the recommender systems community rarely gauges the energy consumption of experiments, making it difficult to estimate their environmental impact.
Consequently, this impedes the recommender systems community in making evidence-based decisions to reduce their environmental impact.

The second reason for measuring and reporting the energy consumption of recommender systems experiments is the ability to compare the energy consumption of running recommender systems.
Like run time and accuracy, energy consumption could be a significant factor in deciding which recommender systems algorithm is best for a given task.
The ability to compare the energy consumption of related approaches would add a precise metric to gauge the efficiency of recommender systems.

To our knowledge, no straightforward solution enables recommender systems practitioners to measure the energy consumption of their experiments easily.
Modifying existing software-based power consumption estimation solutions to work for recommender systems experiments is possible with additional engineering and development effort. 
However, as outlined in the following paragraphs, these solutions would still lack accuracy or usability.
 
Software packages exist to measure energy consumption for general computing~\cite{JMLR:v21:20-312,benoit_courty_2024_11171501}.
However, they have inherent shortcomings~\cite{powermeter}, e.g., the inability to support all hardware configurations and operating systems.
For example, some hardware, e.g., memory and cooling, often do not have sensors, meaning energy consumption must be estimated, making software energy meters inaccurate and infeasible for scientific reproducibility.
Furthermore, system settings may restrict software from reading sensors, and some software is only available for specific operating systems or CPU architectures.
Finally, the energy consumption estimation software may not support infrequently used, modified, and new hardware due to missing interfaces.

Hardware energy meters are often found in commercially available off-the-shelf smart plugs sold for usage in smart home systems.
These enable measuring the energy draw at high frequency, e.g., at least once a second, providing accurate measurements of the energy use of connected devices.
They are also independent of the hardware and software configuration of the computing resource.
While plenty of open software for reading smart plug sensors exists, we cannot find software that can integrate smart plugs from different manufacturers with recommender systems experiments and provide monitoring capabilities.

To fill this gap, we present EMERS, a software library that simplifies measuring, monitoring, recording, and sharing the energy consumption of recommender systems experiments.
EMERS is open-source and available on GitHub\footnote{\url{https://code.isg.beel.org/emers}}.

\section{EMERS}

\begin{figure}[t]
    \centering
    \includegraphics[width=\linewidth]{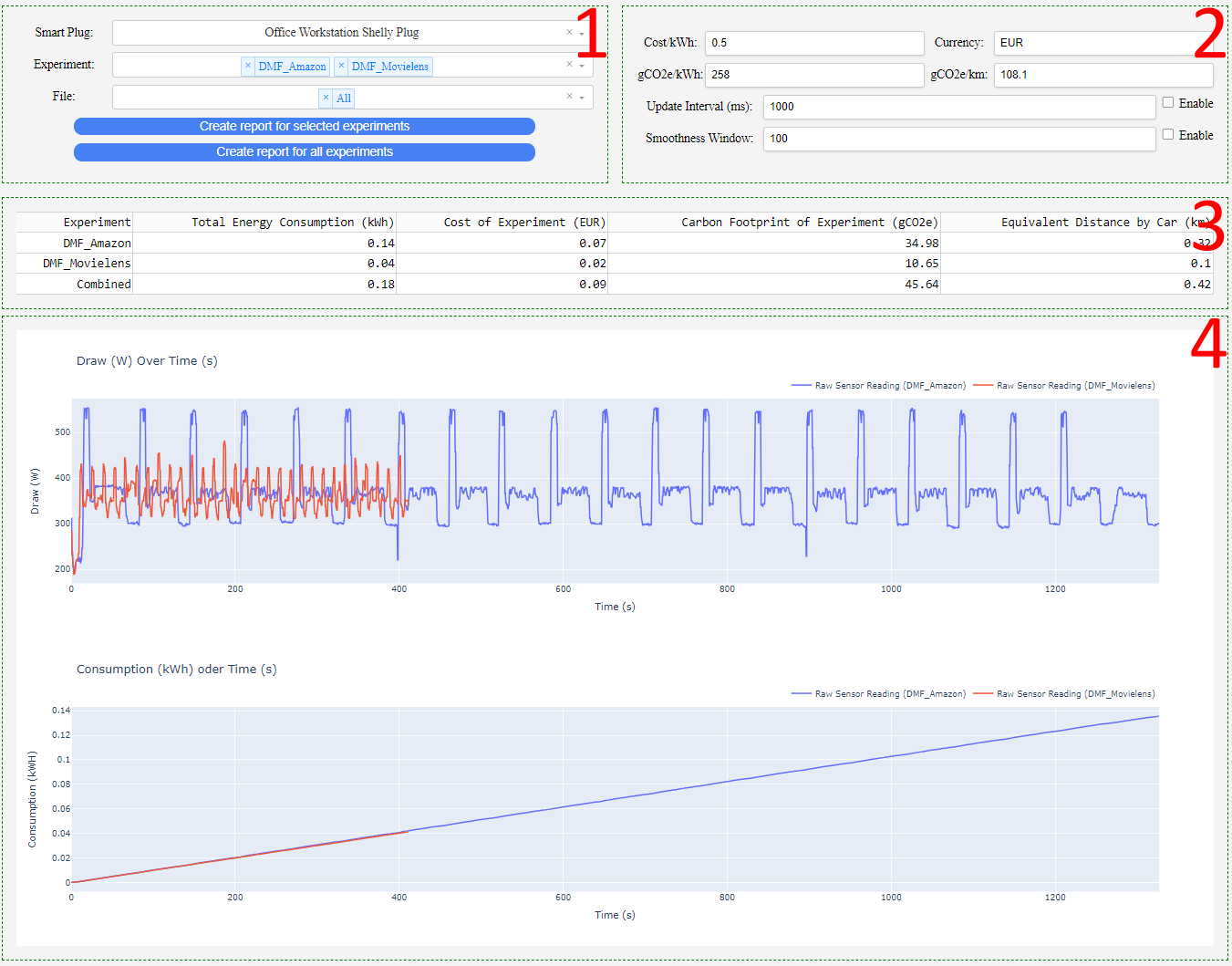}
    \caption{The user interface for EMERS is divided into four regions. We mark each region with a red number. (1) Experiment Selection and Report Generation. (2) Energy Cost, Carbon Footprint, and Graph Settings. (3) Experiment Information. (4) Energy Consumption Graph.}
    \label{fig:ui}
\end{figure}

EMERS is an open-source and platform-independent Python library that simplifies measuring, monitoring, recording, and sharing the energy consumption of recommender systems experiments.
EMERS reads and logs energy measurements from smart plugs, organizes them based on the associated experiment, provides a user interface to monitor and analyze measurements, and creates a standardized, automated report to share with the community.
While smart plugs, or energy meters, come with a small cost in terms of hardware, we believe the accuracy of measurements and compatibility with all hardware systems compensate for it.

\subsection{Features}
EMERS has four features: (i) a user interface for monitoring and analyzing the energy consumption of recommender systems experiments, (ii) standardized energy consumption report generation, (iii) integrated, per-experiment energy consumption logging through our API, and (iv) standalone energy consumption logging with a Python script. 
We provide a detailed description of these features below.

(i) EMERS offers a user interface to interactively monitor, visualize, and analyze the energy consumption of recommender systems experiments.
Figure \ref{fig:ui} shows the user interface with its four regions.
\textbf{Region 1} contains dropdown menus to select which experiments should be monitored.

It also contains buttons to generate the reports detailed in below.
\textbf{Region 2} enables configuring energy consumption and visualization settings.
Notably, the energy cost per kWh and the carbon footprint of energy per kWh can be configured here.
\textbf{Region 3} contains a table that displays energy consumption statistics per experiment and of all selected experiments.
An experiment's energy consumption and associated impact regarding monetary cost and carbon footprint is immediately tangible.
\textbf{Region 4} consists of two graphs, where the first displays the power draw per timestamp, and the second shows the total power draw over time.
The graphs offer live visualization of experiments, enabling practitioners to monitor energy consumption in real time.

(ii) EMERS generates a report of the energy consumption of recommender systems experiments.
The report contains the graphical visualizations and tabular energy consumption data shown in the user interface.
It uses the configuration made in the user interface, e.g., the selected experiments, as well as the cost, carbon footprint, and visualization settings.
The energy consumption report is designed to be added to, e.g., a paper or a document detailing the experiments performed.

(iii) EMERS can measure energy consumption with integrated, per-experiment logging through an API.
For example, recommender systems practitioners can import EMERS into their code to measure the energy consumption of the training process of their deep neural network recommender system with a single API call.
This feature enables targeted energy consumption measurement in experiments and organizing energy consumption logs.
Hereby, practitioners may define the scope of an experiment and decide the granularity by which they would like to measure energy consumption.
For example, an experiment could be defined as a training process of one algorithm or the entire evaluation pipeline of multiple algorithms.

(iv) EMERS offers standalone energy consumption logging.
For example, EMERS can run independently of an experiment or other software to continuously measure and log energy consumption.
Thereby, EMERS runs as a lightweight background process that polls the smart plug with the desired frequency and organizes the measurements in a log file.
Standalone logging with EMERS helps gauge energy consumption for a broader range of use cases over an unspecified time without being tied to specific execution blocks.

\subsection{Requirements}

Measuring energy consumption with EMERS requires a computer connected to the same network as the smart plug.
Besides that, EMERS is lightweight and can be run on the computing resource that executes the experiments without noticeable performance impact.
EMERS can also run on a small, single-board computer, e.g., a Raspberry Pi.

The EMERS monitoring user interface runs on a Flask server and can be configured to be accessible only from the host, the local network, or the internet if desired.
It only requires access to the energy consumption logs and no network access to the smart plug.

Smart power plugs can generally be accessed remotely, e.g., through WiFi or Bluetooth.
EMERS supports two types of WiFi-based smart plugs out of the box: the Shelly Plug Plus S\footnote{\url{https://kb.shelly.cloud/knowledge-base/shelly-plus-plug-s-1}}, and the TP-Link Tapo P115\footnote{\url{https://www.tp-link.com/se/home-networking/smart-plug/tapo-p115/}}. 
Both plugs fit the CEE Type 7 socket in continental European countries and are available in variations for other electrical sockets, such as NEMA sockets in the USA.
EMERS energy consumption logging can be used from any computer within the plug's network and is independent of the computing resource that runs the experiments.
Therefore, physical access to the measured computing resource is required only once to install the smart plug.
Furthermore, support for measuring energy consumption with other remotely accessible smart plugs can be easily implemented through a simple interface in EMERS.

\begin{table}
\caption{The idle energy consumption of different hardware configurations.}  
\centering
\resizebox{\linewidth}{!}{
\begin{tabular}{l|l|l|l|l|l}
\toprule
\makecell[tl]{System Type} & CPU & GPU & \makecell[tl]{RAM\\in GB} & \makecell[tl]{Storage\\in TB} & \makecell[tl]{Idle Energy Consumption\\in Watts} \\
\midrule
\makecell[tl]{Windows 11\\Workstation} & \makecell[tl]{Intel Xeon\\W-2255\\@ 3.70 GHz} & \makecell[tl]{NVIDIA\\GeForce\\RTX 3090} & 256 & 2 & 69.15 $\pm$ 2.45 \\
\midrule
\makecell[tl]{Windows 10\\Workstation} & \makecell[tl]{Intel Core\\i7-6700K \\@ 4.00 GHz} & \makecell[tl]{NVIDIA\\GeForce\\GTX 980 Ti} & 128 & 1 & 80.45 $\pm$ 3.45 \\
\midrule
\makecell[tl]{2022 Mac Studio} & M1 Ultra & M1 Ultra & 64 & 1 & 18.55 $\pm$ 1.55 \\
\midrule
\makecell[tl]{2020 MacBook Pro} & M1 & M1 & 16 & 1 & 12.20 $\pm$ 1.3 \\
\bottomrule
\end{tabular}
}
\label{tab:hardware}
\end{table}

\subsection{Measurement Considerations}

Using a smart plug, EMERS implicitly measures the energy consumption of the complete system to which the smart plug is connected.
Therefore, measurements may be affected by factors other than a recommender systems experiment, e.g., other software running on the system and the system's energy consumption without load.
In Table \ref{tab:hardware}, we show our measurements of the idle energy consumption of four different hardware configurations.
Here, idling means the system only runs necessary, repetitive background tasks with low computing demands.
We observe that the idle energy consumption has a consistent baseline value with minor deviations.

If the system does not run a recommender systems experiment, EMERS can effectively measure this noise with its standalone energy consumption logging mode.
However, other computationally demanding software, especially with varying energy consumption, may significantly impact the measured energy consumption.
Ideally, a system that runs computationally demanding tasks besides the recommender systems experiments should not be used when the goal is measuring energy consumption.
This is not a limitation of EMERS in particular but of energy measurement methods in general.
To illustrate, in theory, even more sophisticated energy measurement methods are challenged by the confounding effects of running multiple software packages on a system, e.g., on cooling and memory, but precisely measuring only the impact of one of them.
Therefore, we recommend running EMERS on an otherwise idling system and measuring the idle energy consumption with EMERS, if desired.

\subsection{Measurement Reliability}

The reliability of measurements with EMERS largely depends on the chosen smart plug.
For example, considering the two smart plugs that EMERS supports out-of-the-box, we find that the Shelly Plug Plus S allows more frequent measurements than the TP-Link Tapo P115, providing a higher resolution.
However, comparing the measurements of both smart plugs with different measurement intervals but on the same hardware configuration and recommender systems experiment, we find differences of $<0.1\%$ over whole experiments.
Therefore, we conclude that the measurements from at least the two aforementioned smart plugs are comparable and reliable due to their similarity.

EMERS receives the measurements from a smart plug's sensors, for example, via HTTP API, and writes them to a text file.
In this sense, EMERS trusts the smart plug and guarantees reliable measurements as long as it retains access to its log file and has a stable connection with the smart plug.

Finally, EMERS notifies the user if any unexpected error or connection issue arises.
The user may then decide whether the measurements are still reliable based on the error. 
Furthermore, the EMERS user interface allows visual inspection of measurements, providing an additional layer for auditing measurements.

\section{Demo}
In the demo video\footnote{\url{https://youtu.be/vmXOcrVpRDg}}, we show the setup of EMERS and present its features.

\emph{Setup}: We demonstrate how to install a smart plug that is required by EMERS and then how to get EMERS and register the smart plug.

\emph{Features}: We provide an overview of the monitoring user interface of EMERS, with its four regions.
Then, we show how EMERS is integrated with recommender systems pipelines to measure energy consumption.
We demonstrate how the EMERS monitoring user interface enables recommender systems practitioners to monitor the energy consumption of their experiments actively.
Finally, we present the energy consumption report that EMERS generates.

\bibliographystyle{splncs04}
\bibliography{sample-base}

\end{document}